# Tunable Magnetic Anisotropy in Patterned SrRuO$_3$ Quantum Structures: Competition between Lattice Anisotropy and Oxygen Octahedral Rotation

*Hongguang Wang$^*$, Gennadii Laskin, Weiwei He, Hans Boschker, Min Yi, Jochen Mannhart &, and Peter A. van Aken*

Dr. H. Wang,  Dr. G. Laskin, Dr. H. Boschker, Prof. J. Mannhart, Prof. P. A. van Aken
Max Planck Institute for Solid State Research, Stuttgart, 70569, Germany
Email: hgwang@fkf.mpg.de
W. He, Prof. M. Yi
State Key Lab of Mechanics and Control of Mechanical Structures & College of Aerospace Engineering, Nanjing University of Aeronautics and Astronautics (NUAA), Nanjing 210016, China

Keywords: Quantum Structures, SrRuO$_3$, Scanning Transmission Electron Microscopy, Magnetic Anisotropy, Lattice Anisotropy, Oxygen Octahedral Rotation

Artificial perovskite-oxide nanostructures possess intriguing magnetic properties due to their tailorable electron-electron interactions, which are extremely sensitive to the oxygen coordination environment. To date, perovskite-oxide nanodots with sizes below 50 nm have rarely been reported. Furthermore, the oxygen octahedral distortion and its relation to magnetic properties in perovskite oxide nanodots remain unexplored yet. Here, we have studied the magnetic anisotropy in patterned SrRuO$_3$ (SRO) nanodots as small as 30 nm while performing atomic-resolution electron microscopy and spectroscopy to directly visualize the constituent elements, in particular oxygen ions. We observe that the magnetic anisotropy and RuO$_6$ octahedra distortion in SRO nanodots are both nanodots' size-dependent but remain unchanged in the first 3-unit-cell interfacial SRO monolayers regardless of the dots' size. Combined with the first principle calculations, we unravel a unique structural mechanism behind the nanodots' size-dependent magnetic anisotropy in SRO nanodots, sugguesting that the competition between lattice anisotropy and oxygen octahedral rotation mediates anisotropic exchange interactions in SRO nanodots. These findings demonstrate a new avenue towards tuning magnetic properties of correlated perovskite oxides and imply that patterned nanodots could be a promising playground for engineering emergent functional behaviors.

1. **Introduction**

Complex oxide heterostructures have a wealth of magnetic states stemming from the specific coupling between spin, charge, orbital, and lattice degrees of freedoms, thereby leading to intriguing applications in emerging spintronic devices[1-3]. Tremendous efforts have been devoted to looking for novel pathways to tune the magnetic anisotropy in oxide heterostructures through controlling electron-electron correlations[4, 5]. On the one hand, one can directly tune the intrinsic magnetocrystalline anisotropy, which derives from the oxygen-coordination environment during material growth[6]. Famous strategies include strain engineering[7], interface engineering,[8] applying chemical[9], or physical pressures[10]. On the other hand, magnetic anisotropy is also tailorable after growth by extrinsic contributions such as applying an electric field[11, 12], and controlling the dimensionality[13]. Materials with reduced dimensionality have altered strain fields and reconstructed electronic structures, offering a playground for feasibly engineering magnetic states[14, 15]. For instance, Wenisch *et al.* demonstrated a local control of magnetic anisotropy of (Ga, Mn)As heterostructures by patterning into nanolines[16]. Furthermore, due to electron-system confinements from 3D, nanodots have modified energy levels, presumably giving rise to unique magnetic behaviors[17, 18]. To date, most reported nanodots are constructed from standard semiconductors[19] or metals[20]. Patterned nanodots comprising correlated perovskite oxides have rarely been studied because of the difficulty of patterning into small scales. In the latest progress, Laskin and Wang *et al.* patterned ferromagnetic perovskites to scales of tens of nanometers and found that the magnetic transition temperature ($T_C$) is controllable by changing the size of patterned nanodots[21]. However, the magnetic anisotropy of nanodots fabricated with correlated oxide materials and the underlying fundamental physical mechanisms still remain open questions.

Oxide-perovskite heterostructures are well known for their strong coupling between magnetic properties and centrally coordinated oxygen octahedra, where the deformation of oxygen octahedra influences the magnetic behavior through metal-oxygen-orbital hybridization and rotations of oxygen octahedra change the spin order by breaking the degeneracy of *d* orbitals[12, 22]. Kan *et al.* demonstrated that the magnetic anisotropy of a correlated ruthenate layer is controllable by an octahedral rotation determined through the atomic-scale design of the oxygen octahedral coupling at the interface[23]. Lee *et al.* observed a reversed magnetic configuration in manganite heterostructures resulting from an octahedral deformation rather than a rotation of oxygen octahedra[24]. Although numerous routes for tuning the magnetic anisotropy have been reported, the underlying structural mechanism concerning the respective role of an octahedral deformation and rotation is still under debate, demanding further clarifications. Furthermore, the role of an interfacial anisotropy and a bulk anisotropy in heterostructures also requires in-depth studies. Compared with most reported strategies, in this work, nanodots are fabricated after growth and have fewer additional side effects on the magnetic behavior, making it easier to understand how deformation and rotation of oxygen octahedra influence nanodots' magnetic properties.

Here, itinerant ferromagnetic $SrRuO_3$ (SRO, bulk Curie temperature $T_C \approx 160$ K) was selected as a model material for fabricating nanodots. It belongs to a family of quasi-cubic perovskite oxides[25]. The growth of a single-crystalline SRO thin film on appropriated substrates, *e.g.,* STO, can be achieved routinely[26, 27]. Structural modulations and magnetic behavior of epitaxial SRO thin films on various substrates have been well-documented[28-30], which helps to understand the the magnetic properties of SRO nanodots. High-quality SRO thin films exhibit a strong magnetocrystalline uniaxial anisotropy, which is highly sensitive to slight variations in stoichiometry, lattice structure and temperatures[31-34]. Another interesting point is that the electronic structure of Ru $4d$ orbitals is controllable by quantum-confinement effects [35, 36].

Here, we investigate the magnetic anisotropy of 5 mm × 5 mm-sized SRO nanodots arrays patterned with electron-beam lithography (EBL) (see Experimental Section for details). The SRO nanodots size down to 15 nm, nearly an order of magnitude smaller than most previously reported results[37]. As the size of the nanodots decreases, its ferromagnetic behavior is investigated, and the magnetic easy-axis is found to rotate from the out-of-plane (OOP) direction towards the in-plane (IP) direction (see Experimental Section for details). Scanning transmission electron microscopy (STEM) results indicate that a structural variation in the bulk part of the SRO layer gives rise to the observed alteration of the magnetic anisotropy. First-principle calculations were performed to evaluate the role of competing mechanisms in terms of deformation and rotation of oxygen octahedra.

2. Results

## 2.1. Microstructure and magnetic behaviors

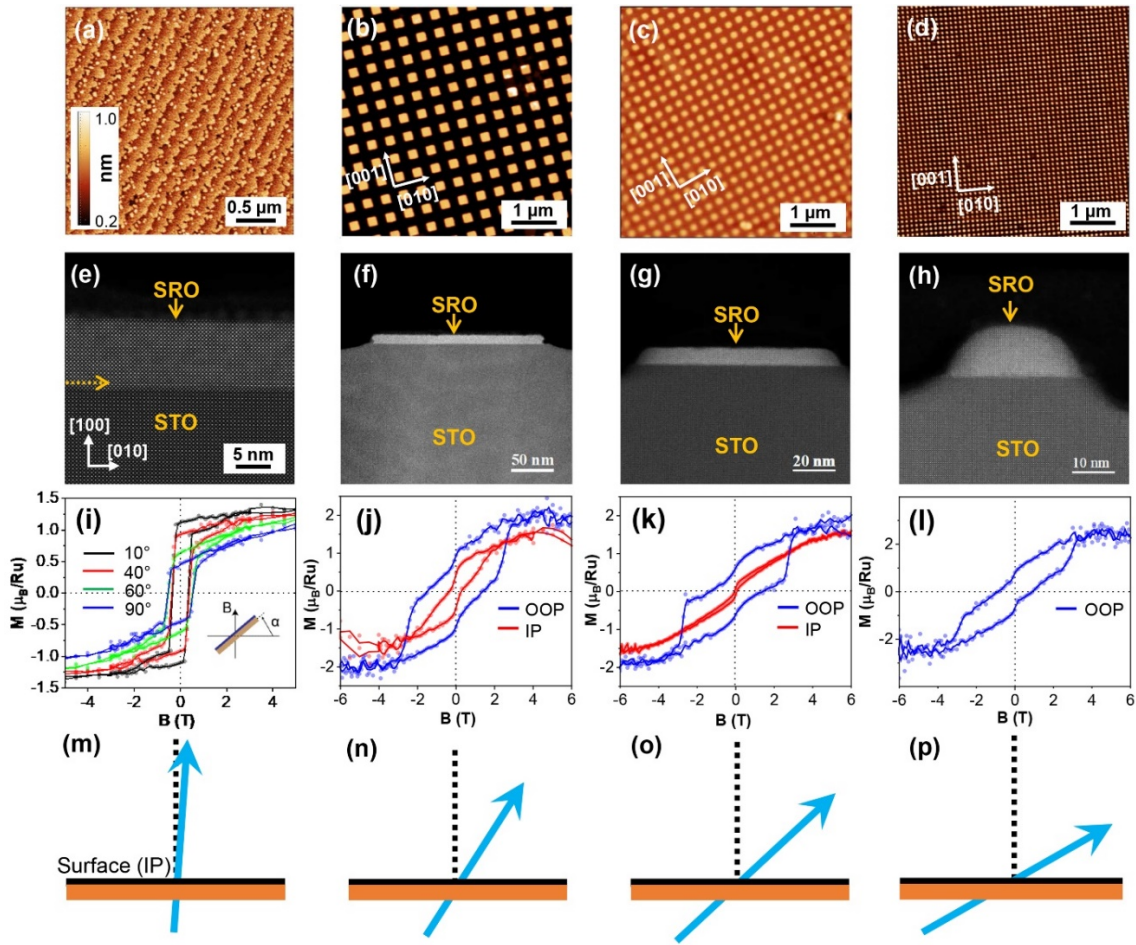

Figure 1. Structure and magnetic properties of the epitaxial SRO layer and patterned SRO nanodots arrays. (a)-(d) show the surface topography of the SRO thin layer and the array of nanodots with a diameter of 200 nm, 80 nm, and 30 nm, respectively. (e)-(h) are corresponding HAADF-STEM images of the cross-sections. (i)-(l) are the corresponding magnetic hysteresis $M(B)$ loop measurements. Magnetic hysteresis $M(B)$ investigations of an epitaxial SRO layer (i) were performed at several substrate orientations and at a temperature of $T = 15$ K. For nanodots, magnetic hysteresis $M(B)$ measurements were conducted at both IP and OOP orientations at $T = 15$ K. (m)-(p) are corresponding schematic diagrams showing the orientation of the magnetic easy-axis at 15 K.

Figure 1(a) shows a topographic atomic force microscopy (AFM) image of an epitaxial SRO thin film right after growth without noticeable surface contamination by hydrocarbons and/or air dust. The terrace structure with the height of about one unit cell of the SRO lattice is well visible, showing a smooth surface of the SRO thin film. It indicates an excellent epitaxial quality of the SRO thin film. Figs. 1(b), (c), (d) correspond to the topography of the array of nanodots with sizes of 200, 80, and 30 nm, respectively. Nanodots are well arranged on the top of STO substrate, and the shape of nanodots deviates from a rectangle to a circle as the size is smaller than 200 nm. The number of nanodots increases with the decrement of the nanodot size. For nanodot arrays with a dot size below 20 nm, the number of SRO nanodots is on/above the order of magnitude of $10^9$. Due to inhomogeneity during patterning, *i.e.*, resist thickness variations and substrate charging, defects may appear in the resulting structure, such as missing or distorted nanodots (Figure 1(d)). This phenomenon becomes more evident for dots as small as 15 nm (see Supplementary Figure S1). The probability of such effects increases with reduced nanodot size and spacing, but typically does not exceed several percent of the total number of nanodots for an array of dots with a size above 30 nm.

Cross-sectional TEM specimens of the nanodots were prepared for observing the structural quality and interface coherency with the substrate. Figure 1(e) displays a high-angle annular dark-field (HAADF) image of the cross-section of a SRO thin film. The SRO thin film is grown on top of STO by perfect layer-by-layer stacking. Since signal intensities of the HAADF image are proportional to the mass of the constituent atoms, the interface between SRO and STO is self-evident. The thin film and the interface are free of any visible defect. STEM results for SRO nanodots in Figs. 1(f,g,h) show that after patterning the samples still keep a high epitaxial quality with a clean interface without any detectable defects, revealing that the crystalline structure remains nearly intact during the patterning process. Due to technical limits, the nanodots' side edge inevitably becomes curved as the nanodot size decreases. The curved shape of the substrate near the nanodot and the shape of the nanodot itself are caused by the non-uniform removal of material during dry etching with Ar ions. During the TEM sample preparation, the nanodot was cut through its center with a lamella thickness of about 20 nm. Therefore, the cross-section provides a projection through almost the entire nanodot.

The magnetic hysteresis loops (*M(B)*) have been measured to characterize the ferromagnetic properties of materials (see the Experimental Section for details). The overall shape of the hysteresis loops reflects the materials' magnetic anisotropy. If the angle between the easy axis and the direction of the applied magnetic field is small, the remanent magnetization $M_{rem}$ tends to be close to the saturation magnetization $M_{sat}$, yielding a rectangular-like shape of the hysteresis loop. Figure 1(i) shows measured hysteresis loops as the unpatterned SRO thin film rotates around the [001] axis relative to the magnetic field *(B)*. Four different orientations (10°, 40°, 60°, 90°) of the substrate were applied for the measurement of *M(B)* curves (See Supplementary Figure S2 for the 0° results of SRO grown with similar conditions). The loop shapes deviate from a rectangular shape as the magnetic easy-axis direction rotates away from the applied magnetic-field direction. The shape at a tilt angle of 10° is more "rectangular-like" than for larger tilt angles, indicating that the easy axis is almost parallel to the film normal. For a uniaxial magnetic system like SRO, the sample orientation with respect to the applied magnetic field direction yielding the largest remanent-to-saturation-magnetization $M_{rem}/M_{sat}$ ratio indicates the magnetic easy-axis. We observe that the ratio decreases from $(M_{rem}/M_{sat})_\perp = 0.85$ to $(M_{rem}/M_{sat})_{//} = 0.34$ with an increasing tilt angle from $\alpha = 10°$ to $\alpha = 90°$, indicating that the epitaxial SRO thin film has an uniaxial perpendicular magnetic anisotropy. Figs. 1(j, k, l) correspond to magnetic hysteresis *M(B)* loops of an array of 200 nm, 80 nm, and 30 nm-sized nanodots patterned from an as-grown SRO thin film, measured in OOP and IP orientations, respectively. For the hysteresis loop of the 200 nm-sized nanodots array (Figure 1(j)), $M_{rem}/M_{sat}$ changes from $(M_{rem}/M_{sat})_\perp = 0.51$ in the OOP to $(M_{rem}/M_{sat})_{//} = 0.32$ in the IP orientation. For the hysteresis loop of the 80 nm-sized nanodots array (Figure 1(k)), we measure $(M_{rem}/M_{sat})_\perp = 0.44$ and $(M_{rem}/M_{sat})_{//} = 0.12$, which reduces further for nanodots down to 30 nm (Figure 1(l)), yielding a much weaker magnetic signal of the nanodots. For the 30 nm-sized nanodots, only the hysteresis loop in the OOP direction was obtained reproducible with $(M_{rem}/M_{sat})_\perp = 0.39$. To calculate these ratios, the remanent magnetization was determined by linear extrapolation of the magnetization at high magnetic fields equal 0. Since the measurement uncertainty of caluculated ratios is relatively large and hard to be precisely determined, the presented ration numbers are used to guide the changing trend of magnetic anisotropy rather than definitive values. The $(M_{rem}/M_{sat})_\perp$ drops gradually as the nanodots' size decreases, revealing that the magnetic easy-axis rotates from the OOP direction toward the IP direction, which is sketched in Figs. 1(m, n, o, p) for an SRO thin film, 200 nm-sized nanodots, 80 nm-sized nanodots, and 30 nm-sized nanodots, respectively. There is a step at small fields in both IP and OOP loops, which may originate from several mechanisms. Nevertheless, the contribution of magnetic impurity can be fully excluded in this work (see Experimental Section for detailed explanations). Magnetic hysteresis loops as a function of the lateral spacing between nanodots suggest that the magnetic interactions within and between nanodots are likely causes (see

Supplementary Figure S3)

## 2.2. Lattice structure and local chemistry at the heterointerface

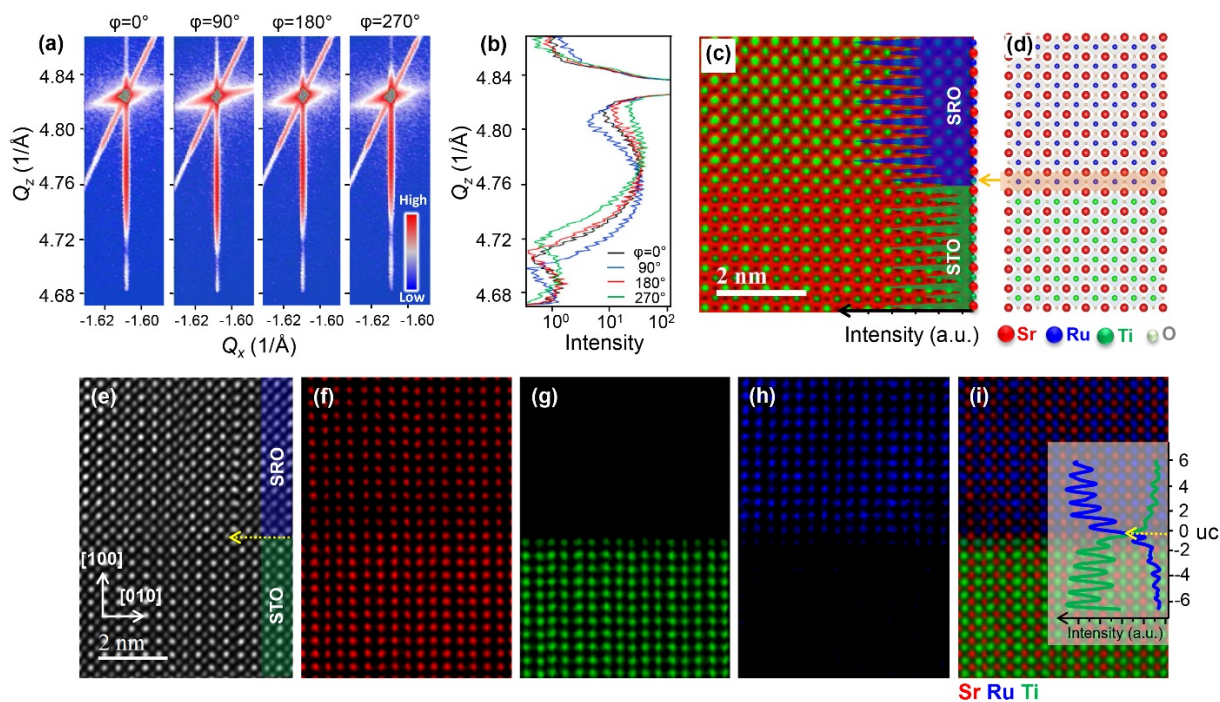

Figure 2. Lattice structure and local chemistry at the heterointerface between SRO and STO. (a) RSMs of an SRO thin film measured around the ($\bar{1}$03) peak of STO at four different $\varphi$ angle orientations of the substrate, $\varphi = 0°$, 90°, 180°, and 270°. (b) Line-by-line averaged intensity profiles along the $Q_z$ direction for different $\varphi$ angles, $\varphi = 0°$ (black), 90° (blue), 180° (red), and 270° (green). (c) Overlay of simultaneously acquired HAADF and ABF images of the thin film. The inset at the right-hand side is the laterally averaged intensity profile of the HAADF signal. The yellow arrow indicates the interfacial B(Ru/Ti)-O layer. d) The corresponding schematic diagram of (c). (e)-(i) STEM-EELS elemental mapping at the interface of a 30 nm-sized SRO nanodot on a STO substrate. (e) HAADF-STEM imaging of a region close to the interface. (f) Sr elemental map using the Sr-$L_{2,3}$ edges. (g) Ti elemental map using the Ti-$L_{2,3}$ edges. (h) Ru elemental map using the Ru-$L_{2,3}$ edges. (i) The composite map using the color scheme from (f)-(h). The inset shows the signal intensity profile of Ti and Ru at the interfacial region.

In order to determine the structure of the epitaxial SRO thin films, we have applied X-ray diffraction (XRD) techniques. Figure 2(a) shows four reciprocal space maps (RSMs) of a SRO thin film at different angles $\varphi$ of the STO substrate with respect to the incoming X-ray beam, proving that the epitaxial SRO thin film is untwinned. The intense central maximum in the upper part of the RSMs corresponds to the ($\bar{1}$03) STO diffraction peak, and the vertically elongated peaks correspond to SRO. These peaks are shifted to smaller $Q_z$, indicating a larger OOP lattice parameter in SRO than in STO. The IP lattice parameters along $Q_x$ are identical, confirming that the epitaxial SRO thin film is fully strained by the STO substrate (see Supplementary Figure S4 for further XRD data). Figure 2(b) shows the line-by-line averaged intensity profiles along the $Q_z$ direction extracted from the RSM maps in Figure 2(a). It shows that the SRO peak positions are shifted in the $Q_z$ direction by a small amount, which depends on the angle $\varphi$, revealing that SRO has a distorted orthorhombic structure. The profiles for the SRO peak along $\varphi = 0°$ and $\varphi = 180°$ coincide, implying that an orthorhombic distortion cannot be observed in these directions. In contrast, the peaks at $\varphi = 90°$ and $\varphi = 270°$ are shifted relative to the central ones ($\varphi = 0°$ and $\varphi = 180°$), allowing to determine the degree of orthorhombicity. The orthorhombic unit cell can be approximated as a tilted pseudo-cubic unit cell with a tilt angle $\alpha = \arctan(\Delta Q_z / \Delta Q_x)$, where $\Delta Q_z$ and $\Delta Q_x$ are distances in reciprocal space along $x$ and $z$ between SRO peaks for different $\varphi$ angles. From the $\varphi = 90°$ and $\varphi = 270°$ peaks, a tilt angle of $\alpha = 0.49°$ has been determined, which is in good agreement with literature data.[38, 39] RSM studies of the nanodots arrays were also performed (see Supplementary Figure S5), but due to low counting statistics from the small number of patterned material we cannot draw solid conclusions about the material structure. Therefore, we characterized the structure by atomically resolved STEM techniques.

Figure 2(c) shows a superimposed image of a simultaneously acquired annular bright-field (ABF) and HAADF image of the interface region between SRO and STO. All constituent elements, including oxygen, are clearly resolved. The corresponding sketch of the interfacial structure is displayed in Figure 2(d). The oxygen octahedra at both sides of the interface are well connected by corner-sharing. The horizontally averaged HAADF intensity profile in the inset of Figure 2(c) depicts that the intensity of the first "Ru-O monolayer" is composed of a mixed Ru-Ti-O layer, which is about one unit cell thick. It is worth noting that the HAADF signal profiles are identical for SRO thin films and patterned nanodots on STO substrates. Figures 2(e-i) show STEM-EELS results of the elemental distribution of Sr, Ti, Ru at the interface between SRO and STO for a 30 nm-sized SRO nanodot at the atomic scale. The Sr-$L_{2,3}$ map in Figure 2(f) reveals a homogeneous Sr distribution across the imaged area. The Ti-$L_{2,3}$ (Figure 2(g)) and Ru-$L_{2,3}$ (Figure 2(h)) intensity maps indicate the different regions of STO and SRO, respectively. The composite image (Figure 2(i)) of the Sr, Ti, and Ru maps directly visualizes the elemental distribution around the interface. The Ti and Ru signal intensities at the interface (c.f. inset of Figure 2(i)) confirm a weak intermixing of Ti and Ru within about one unit cell on both sides of the interface, yielding the same chemical distribution as at the interface of SRO thin films on STO substrates. Similar results have been reported at the interface of SRO-STO superlattices, which were grown by pulsed laser deposition (PLD)[29].

**2.3. Lattice anisotropy and oxygen octahedral rotation**

Quantitative STEM analysis has been performed to investigate alterations of the lattice anisotropy ($c/a$) and octahedral rotations $\theta = \pi - \alpha$, which are critical factors for influencing the magnetocrystalline anisotropy of SRO nanodots. The $c/a$ values were determined from HAADF images, whereas for the octahedral rotation angles we analyzed ABF images recorded simultaneously. It is worth noting that, since ABF results were acquired in the cross-section orientation, we exclusively have measured amplitudes of rotation angles along the in-plane rotation axis. The 2D $c/a$ maps of the unpatterned SRO thin film, the 80 nm-sized nanodots, and the 30 nm-sized nanodots are displayed in Figure 3(a), (b), and (c), respectively, using the same color scale. In the SRO part of the unpatterned thin film, we observe an even distribution of the $c/a$ values over the entire SRO layer. In contrast, the SRO part of patterned nanodots displays a maximum of $c/a$ values close to the interface, decreasing while moving toward the surface. The averaged $c/a$ value in the SRO layer becomes smaller as the size of the nanodots reduces (see Supplementary Figure S6). To precisely determine the $c/a$ values in the nanodots, we fit Gaussian functions to atomically resolved STEM HAADF images and determine the coordinates of the constituent elements. Figure 3(d), (e), (f) correspond to $c/a$ values of the first 15 unit cells of the unpatterned SRO thin film, an 80 nm-sized nanodot, and a 30 nm-sized nanodot. The calculated $c/a$ values using interatomic distances of the A- (Sr) and B-site (Ru/Ti) atoms match well with each other for all studied samples. Comparing these values to an unpatterned SRO thin film with a constant $c/a$ value of $1.015 \pm 0.005$, the $c/a$ values of patterned nanodots reach a maximum close to the SRO/STO interface and drop steeply as moving away from the interface. Most importantly, we can observe a clear trend that the smaller the size of the nanodot, the smaller the $c/a$ value at the same distance from the interface, demonstrating that the overall degree of lattice anisotropy is reduced for nanodots with smaller sizes. Nevertheless, the $c/a$ values of the first 3 SRO monolayers remain almost unchanged for all nanodots sizes.

Based on the fitted coordinates of the oxygen positions, Figure 3(g, h, i) presents the IP ($\theta_1 = \pi - \alpha_1$) and OOP ($\theta_2 = \pi - \alpha_2$) angles between between neighboring $RuO_6$ octahedra for the first 15 unit cells of an unpatterned SRO thin film, an 80 nm-sized nanodot, and a 30 nm-sized nanodot. For the unpatterned SRO thin film, the angles $\theta_1$ and $\theta_2$ remain unchanged at a value of 0°, since the STO substrate fully strains the SRO lattice. The rotation of the oxygen octahedra appears in SRO nanodots due to the patterning. For smaller nanodots, the SRO lattice has a higher degree of freedom for strain relaxation, deviating the rotation angles of $RuO_6$ octahedra more from 0°. For 80 nm-sized nanodots, the rotation angles show a slight increase to around 2°. In contrast, the octahedral rotations in 30 nm-sized nanodots are more prominent with an angle of approximately 15°, which is similar to bulk SRO. As moving away from the interface, the OOP plane rotation occurs earlier than the IP rotation due to a gradual IP strain relaxation. Interestingly, the $RuO_6$ octahedra of the first 3 unit cells exhibit identical results on rotation angles with a value of 0° for both $\theta_1$ and $\theta_2$. It implies that the crystal structure of the fully strained first 3-unit-cell layers is too robust to be modified by the patterning process.

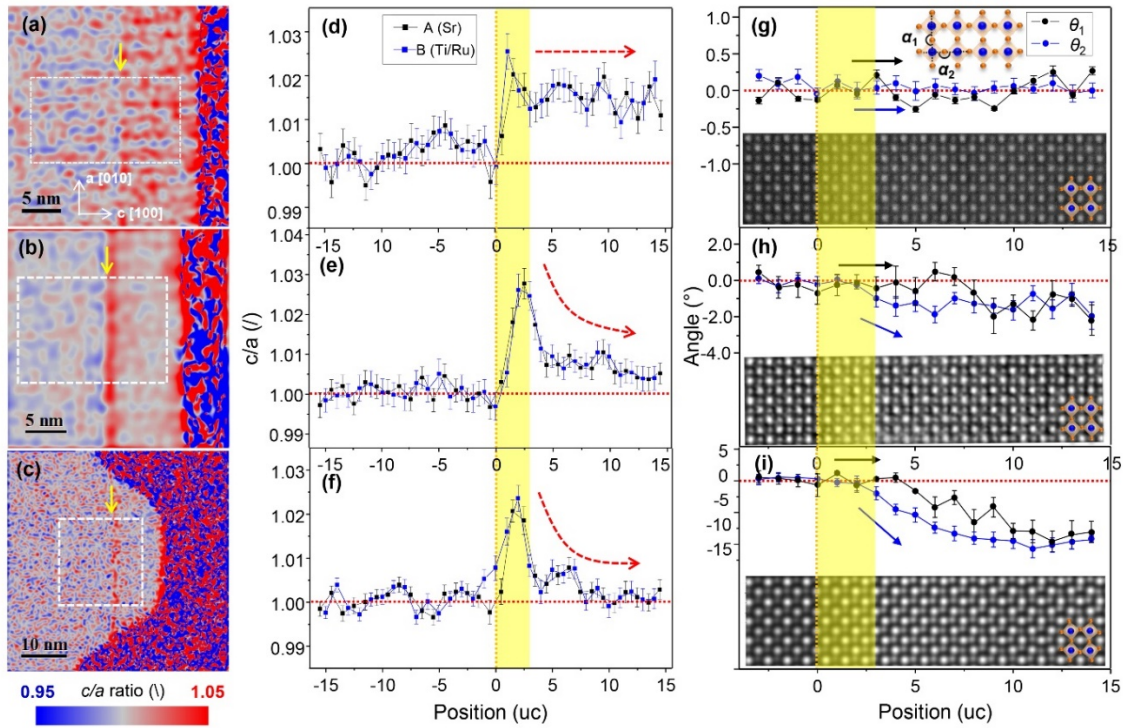

Figure 3. Lattice anisotropy and oxygen octahedral rotation in SRO. (a), (b), (c) are the 2D *c/a* maps of the unpatterned SRO thin film, 80 nm-sized nanodots, 30 nm-sized nanodots, respectively. The vertical arrows denote the interface between SRO and STO. (d), (e), (f) correspond to the vertically-averaged *c/a* values of the regions marked by boxes. (g), (h), (i) depict the relating rotation angles ($\theta = \pi - \alpha$) in the IP ($\theta_1$) and OOP directions ($\theta_2$) in the boxed areas in (a), (b), and (c). The definition of $\alpha_1$ and $\alpha_2$ is shown in the inset of (g). The rotation angle $\theta$ denotes a 2D projection of a 3D bonding angle determined from the oxygen positions in the inverted ABF images below the plots. Position 0 represents the surface TiO layer of the STO substrate.

## 2.4. First-principles calculations

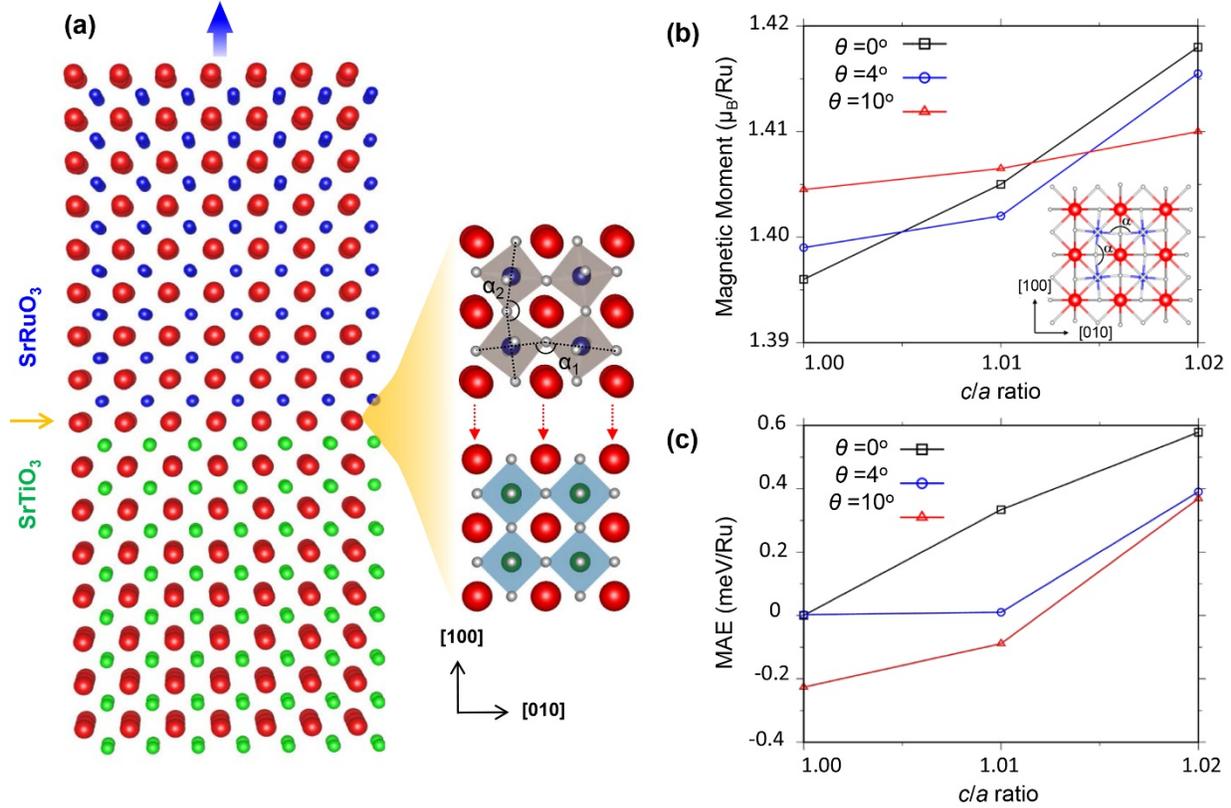

Figure 4. First-principles calculations of the magnetic moment and the magneto-crystalline anisotropy energy (MAE) of Ru in SRO. (a) Sketch of the SRO/STO heterostructure (left panel) and epitaxial growth of pseudo-cubic SRO on cubic STO (right panel). $\alpha_1$ and $\alpha_2$ represent the IP and OOP rotation angles of an oxygen octahedron in a 2×2×2 SRO supercell, respectively. (b) The effect of the $c/a$ ratio and the oxygen octahedra rotation angle $\theta$ on Ru's magnetic moment in SRO. When the oxygen octahedron rotates by angle $\theta$, the bond angle $\alpha$ of Ru-O-Ru satisfies $\theta = \pi - \alpha$. (c) The effect of the $c/a$ ratio and $\theta$ on the MAE of Ru in SRO.

To probe the physical origin of the nanodots' magnetic anisotropy, the magnetic moment and the magnetocrystalline anisotropy energy (MAE) of Ru in SRO were calculated within a 2×2×2 supercell (*i.e.*, $Sr_8Ru_8O_{24}$) using VASP (Vienna *Ab-initio* Simulation Package), which is based on the density-functional theory (DFT). The influences of the lattice anisotropy ($c/a$) and the oxygen octahedral rotation are examined according to the experimentally measured values in Figure 3. Considering the experimental data and previously reported calculations on the SRO system[40-42], an effective Hubbard parameter $U_{eff} = 3$ eV is adopted for our calculations. Here, we define the total energy difference by considering the spin-quantization axis is IP ($E_{//}$) and OOP ($E_\perp$) as MAE, i.e., MAE = $E_{//}$ - $E_\perp$. Thus, a positive (negative) value of MAE indicates the OOP (IP) to be the easy axis.

Figs. 4(b) and 4(c) present the first-principles calculation results, where the magnetic moment and the MAE of Ru are shown for different c/a ratios (c/a = 1.0, 1.01, 1.02) and various oxygen octahedral rotation angles ($\theta$ = 0º, 4º, 10º). The magnetic moment per Ru atom (Figure 4(b)) slightly increases with increasing c/a ratio and is also subtly influenced by $\theta$. Nevertheless, the magnetic moment remains around 1.4 $\mu_B$ per Ru atom irrespective of the c/a ratio and $\theta$. In contrast, the MAE (Figure 4(c)) is highly dependent on the c/a ratio and $\theta$. The MAE is 0 for c/a = 1 and no octahedral rotation ($\theta$ = 0º). Once a strain is applied along the c axis (c/a > 1), the MAE increases with c/a and reaches 0.6 meV/Ru for c/a = 1.02 and $\theta$ = 0º. However, the MAE decreases with an increase of $\theta$. An oxygen octahedron rotation of 10º even leads to an IP easy axis, i.e., a negative MAE, for c/a ≤ 1.01. These results indicate the competitive role of the c/a ratio and $\theta$ on the MAE, i.e., a high c/a ratio increases the OOP MAE, while a large $\theta$ decreases it. Therefore, both the smaller c/a ratio and the large $\theta$ in 30 nm-sized nanodots (Figure 3f, i) synergetically reduce the OOP MAE more, compared to the larger c/a ratio and smaller $\theta$ in 80 nm-sized nanodots (Figure 3e, h). This explains the larger deviation of the easy axis from the OOP direction in 30 nm-sized nanodots.

## 3. Discussion

The magnetization results have unambiguously demonstrated the tunable magnetic anisotropy in patterned SRO nanodot arrays. Since the magnetocrystalline anisotropy in SRO plays a huge role in its overall magnetic anisotropy[25], the effects of surface and shape anisotropy are insignificant here. We attribute the variable magnetic anisotropy of fabricated nanodot arrays to the altered intrinsic magnetocrystalline anisotropy. Generally, the effective uniaxial anisotropy energy $K_\mu$ in a heterostructure can be described as the following relation[43, 44]

$$K_\mu = K_v + K_i/t ,$$

where $K_v$ and $K_i/t$ correspond to the volume anisotropy energy and the contribution from the interface in a ferromagnetic layer with a thickness t. For as-grown SRO thin films with perpendicular uniaxial magnetic anisotropy, the contributions from the volume magnetocrystalline anisotropy $K_v$ and the interfacial crystalline anisotropy $K_i/t$ can not be decoupled, because STEM results reveal that the lattice structure controlling the magnetocrystalline anisotropy remains unchanged throughout the SRO layer. Nevertheless, the role of $K_v$ and $K_i/t$ can be rationally separated for differently sized SRO nanodots. STEM studies identify a 3-unit-cell thick interfacial SRO layer, whose lattice anisotropy and octahedral rotations are stabilized by the symmetry mismatch[45], and show constant values regardless of the nanodots size. In this scenario, $K_i/t$ does not contribute to the magnetic anisotropy in SRO nanodots. In contrast to the interfacial contribution $K_i/t$, we observe a gradual variation of the volume anisotropy energy $K_v$ closely related to the size of SRO nanodots. Octahedral distortions in SRO nanodots are gradually altered while moving away from the interface. Furthermore, the degree of octahedral distortions at the same distance from the interface increases as the size of the SRO nanodots reduces, because the IP biaxial strain relief is more prominent in smaller nanodots. It unveils that $K_v$ plays a dominating role in tuning the magnetic anisotropy of patterned SRO nanodot arrays. Although the cation/anion stoichiometry in perovskite oxides generally affects the $RuO_6$ octahedral rotation [27], its effect can be excluded here, because no apparent changes of the composition of the $SrRuO_3$ can be observed after nanopatterning via energy-dispersive X-ray spectroscopy (EDS) studies. (See Supplementary Figure S7and S8)

The influence of magnetocrystalline anisotropy concerning lattice anisotropy and oxygen octahedral rotations is generally a paradoxon in perovskite heterostructures. $Ru^{4+}$ ions in SRO are usually considered to be in the low-spin state.[46] In fully-strained SRO/STO heterostructures, the $RuO_6$ octahedron is IP compressed, inducing an increased tetragonality. On the one hand, the $RuO_6$ crystal field is modified, enhancing the energy of the lateral $4d$ orbitals and the occupancy of the perpendicular $4d$ orbitals. On the other hand, the orbital hybridization between Ru $4d$ and O $2p$ states is weakened along the $c$ axis, reducing the exchange interaction in this direction. Obviously, the $RuO_6$ crystal-field effect is dominative, which gives rise to a perpendicular magnetic anisotropy in the SRO thin film (Figure 3d, g). In contrast, the structural origin of the magnetic anisotropy in SRO nanodots is size-dependent. When patterning the SRO thin film to nanodots with a relatively large size, the IP strain relaxation leads to a noticeable change of the lattice anisotropy but usually negligible oxygen octahedral rotations. For instance, 80 nm-sized nanodots show an apparent decrease of the lattice anisotropy and minimal octahedral rotations of about 2º in the SRO layer (Figure 3e, h). Hence, the lattice-anisotropy-induced $RuO_6$ crystal-field effects dominate the magnetic anisotropy of large-sized SRO nanodots. However, oxygen octahedral rotations are crucial in controlling the magnetic anisotropy for nanodots as small as 30 nm. We found that the Ru-O-Ru rotation angles change from 2º for 80 nm-sized nanodots to 15 º for 30 nm-sized nanodots (Figure 3i). At the same time, only a slight change of the lattice anisotropy is observed (Figure 3f). DFT calculations have also explained this mechanism, implying that tuning octahedral rotation angles in heterostructures with a minor lattice mismatch is a practical pathway to engineering magnetic behaviors.

We have focused so far on the differences in the lattice structure and magnetic anisotropy as a function of the nanodot size. Interestingly, SRO nanodots also have a complicated lattice structure that evolves with increasing distance from the interface. The first atomic layers close to the interface have a lattice structure with a large c/a ratio, whereas the lattice relaxes further away from the interface, and the *c/a* ratio decreases. SRO is a unique ferromagnet with a large ratio of the magnetocrystalline anisotropy energy and the exchange energy[25]. This implies that the direction of the magnetization can change at very small length scales[47]. Therefore, it can be expected that the nanodots have a complicated magnetic structure, where the magnetization follows the local magnetocrystalline anisotropy energy landscape. Thus, we expect the magnetization within a single nanodot to be oriented relatively out-of-plane close to the interface and more in-plane at the top of nanodots. The unique magnetic properties of SRO in combination with the range of crystal structure variation possible within nanodots makes these nanodots an intriguing playground to engineer novel magnetic configurations for potential applications. Furthermore, in principle, one can design the magnetic behavior in a desired way via nanopatterning when building nanodevices with oxide heterostructures. Via the maintained interfacial magnetic anisotropy, one can also build interface-based nanodevices, in which the magnetic properties are stable during the nanopatterning process.

## 4. Conclusion

This work presents a flexible control of the SRO layer's magnetic anisotropy from the OOP direction to the film plane by patterning the film to SRO nanodots with different sizes. STEM results demonstrate that the magnetic anisotropy modulation in SRO nanodots makes use of controlling volume anisotropy rather than the interfacial magnetocrystalline anisotropy confined within the first 3 unit cell SRO monolayer. The roles of lattice anisotropy and oxygen octahedral rotations on the magnetic anisotropy of the SRO nanodot array are size-dependent, which is revealed by combining STEM studies and DFT calculations. The method presented is not limited to SRO/STO heterostructures, but can be extended to a broad range of heterostructures. One can design the epitaxial layer's magnetic order by selecting a proper substrate and then engineer the magnetic states by patterning them into an array of nanodots with different sizes. Controlling the shape of nanodots is also an exciting topic, which may lead to emergent magnetic phases, *e.g.,* vortex states or skyrmion phases.

## 5. Experimental Section

*Epitaxial growth of SRO thin films*. Epitaxial SRO thin films were grown on $TiO_2$-terminated (100)-oriented STO substrates (Shinkosha) by PLD using a stoichiometric, polycrystalline target (Lesker). The growth temperature was 680 °C, the oxygen pressure was 0.08 mbar, the laser fluence was 2.5 J cm$^2$, the laser frequency was 1 Hz, and the growth rate was 0.05 monolayer per pulse. The growth mode is a layer-by-layer growth as evidenced by the observation of corresponding RHEED oscillations.

*Fabrication of SRO artificial atoms*. The SRO nanodots were fabricated by patterning as-grown epitaxial SRO thin films with electron beam lithography (EBL) in a class-100 (ISO 5) cleanroom, which enables us to make custom shapes by irradiating a suitably sensitive resist material with a focused beam of electrons. The resist material used in this work is poly(*α*-methyl styrene-co-*α*-chloroacrylate methylester) (CSAR, Allresist GmbH, Germany ), which provides superior resolution and requires a smaller exposure dose than the commonly-used resist poly(methyl methacrylate) (PMMA). Electron beam exposure was performed with a design layout using a 100 kV electron beam lithography system (JBX6300, JEOL Co. Ltd.). A hard mask of a thin layer of amorphous $Al_2O_3$ was then deposited on the top by PLD. Afterwards, the resist layer was lifted off by chemically dissolving it in a remover solution, leaving the only hard mask on the top of the SRO layer. The SRO nanodots were finally obtained by dry etching with a flux of ionized argon to pattern the SRO layer (see Supplementary Figure S9 for the fabrication process of SRO nanodots).

*Magnetic measurements.* The magnetic properties of the samples were measured using a Quantum Design MPMS SQUID magnetometer equipped with the reciprocating sample option (RSO) head. The most frequently used measurement parameters were: 4 cm of oscillation amplitude with a frequency of 0.5$^{-1}$ Hz and 10 oscillation cycles per data point. The typical sensitivity of such a measurement is on the order of $10^{-12}$ - $10^{-11}$ Am$^2$ ($10^{-9}$-$10^{-8}$ emu), depending on the experimental conditions. The preparation for the magnetic measurements was done using non-magnetic tweezers and without contact with any magnetic substance. Sincee the patterned nanostructure on 5 mm x 5 mm substrates fit into the setup of our SQUID measurements, no cutting was applied to our samples. The M(H) loops of a bare STO substrate show paramagnetic flat lines without features, proving that no magnetic impurities exist in the initial STO substrates (see Supplementary Figure S10). For all samples, the thickness of the SrRuO$_3$ film is equal to 12 nm (*i.e.*, 32 unit cells of SrRuO$_3$). Bare Every sample was measured twice at different steps of the fabrication process: directly after the film growth and then after being patterned into nanodots. The curves were measured in the range of (-6 T, +6T) with a step size of 0.1 T. For further data analysis, the linear paramagnetic contribution from the STO substrate was subtracted first. The magnetization was then normalized to the estimated total number of Ru atoms present in the nanodots. From large area SEM and AFM scans, the coverage of the substrate was estimated. The thickness and shape of the nanodots were inferred from the TEM data. Combining these data allowed us to calculate the approximate total number of Ru atoms on the substrate. Finally, the magnetic signal has been normalized to the estimated amount of Ru.

*XRD measurements.* XRD data have been measured in-house on the Empyrean set-up (Cu Kα line, $\lambda$ = 1.541 Å) and at the Max Planck Institute beam line which is a part of the synchrotron radiation facility ANKA (Angströmquelle Karlsruhe, KIT, Germany, $\lambda$ = 1.541 Å). The step size for $Q_x$ and $Q_z$ in acquired RSM maps were 0.001 1/Å.

*STEM investigations.* STEM specimens were prepared by focused ion beam (FIB) followed by NanoMill at liquid-N$_2$ temperature. The way, the central part of SRO nanodots was prepared, enables us to perform STEM investigations with high precision.[48] STEM studies were conducted using a spherical aberration-corrected STEM (JEM-ARM200F, JEOL Co. Ltd.) equipped with a cold field emission gun and a DCOR probe Cs-corrector (CEOS GmbH) operated at 200 kV. The STEM images were obtained by JEOL ADF and BF detectors with a convergent semi-angle of 20.4 mrad. The corresponding collection semi-angles for HAADF imaging were 70−300 mrad and for ABF imaging 11−22 mrad. In order to make precise measurements of lattice constants and octahedral rotation angles, ten serial frames were acquired with a short dwell time (2 μs/pixel), aligned, and added afterwards to improve the signal-to-noise ratio (SNR) and minimize the image distortion of HAADF and ABF images. EELS acquisition was performed by a Gatan GIF Quantum ERS imaging filter with dual-EELS acquisition capability with a convergent semi-angle of 20.4 mrad and a collection semi-angle of 111 mrad. Dual EELS spectrum imaging was applied with the dispersion of 1 eV/channel by a 2048 pixel wide detector for simultaneously acquiring the spectrum image of the Sr-$L_{2,3}$, Ti-$L_{2,3}$, and Ru-$L_{2,3}$ edges. The raw spectrum image data were denoised by applying a principal component analysis (PCA) with the multivariate statistical analysis (MSA) plugin (HREM Research Inc.) in Gatan DigitalMicrograph and then smoothed using a spatial filter in Gatan DigitalMicrograph. *c/a* maps were calculated with the lattice-deformation results obtained using a Geometric Phase Analysis (GPA)[49]. The *c/a* value for each unit cell was measured with the Matlab-based image-quantification tool StatSTEM[50]. The Oxygen Octahedra Picker tool was employed for measuring the octahedral rotation angles in IP and OOP directions[51] [52].

*First principle calculations*. The core-electron interaction is described by the projector augmented wave potential, in which the Sr ($4s^2 4p^6 5s^2$), Ru ($4d^7 5s^1$), and O ($2s^2 2p^4$) shells are taken as valence electrons. Owing to the epitaxial growth of the thin SRO film on $SrTiO_3$ (STO) substrate, the in-plane lattice parameters of the epitaxial SRO film are determined by the experimental value (*a*=*b*=3.905 Å) of the bulk STO system[53]. The cut-off energy is set to 500 eV and the convergence criteria for the total energy is $5 \times 10^{-5}$ eV for self-consistent calculations. Moreover, the influence of the oxygen octahedral rotation is studied using a 2×2×2 SRO super cell (*i.e.*, $Sr_8 Ru_8 O_{24}$). The three-dimensional first Brillouin zone is sampled with a 10×10×10 and 5×5×5 Monkhorst-Pack *k*-point mesh for a five-atom unit cell and a fourty-atom super cell, respectively. The local density approximation (LDA) is adopted for the exchange-correlation function. In addition, to account for the strong correlations in *d*-orbitals of Ru, LDA+U with rotationally invariant approach introduced by Dudarev *et al.* is applied[54]. In this way, the effective Hubbard parameter ($U_{eff} = U - J$) is used, where *U* and *J* are the strength of the effective on-site Coulomb energy and exchange interactions of Ru 4*d* states, respectively. The MAE was calculated by the energy difference corresponding to magnetization along different crystal axes. In detail, a collinear spin-polarized self-consistent calculation is performed to obtain the converged charge densities. Then by reading these charge densities, non-collinear non-self-consistent calculations with spin-orbit coupling are carried out to obtain the total energy as a function of spin-quantization axes.

**Supporting Information**

Supporting Information is available from the Wiley Online Library or from the author.

**Acknowledgments**

This project has received funding from the European Union's Horizon 2020 research and innovation programme under Grant Agreement No. 823717 − ESTEEM3. HGW thanks the Max Planck Society for financial support and the support from the Stuttgart Center for Electron Microscopy (StEM) at the Max Planck Institute for Solid State Research. MY acknowledges the support from the 15th Thousand Youth Talents Program of China, German Science Foundation (DFG YI 165/1-1), NSFC (11902150), the Research Fund of State Key Laboratory of Mechanics and Control of Mechanical Structures (MCMS-I-0419G01), and the access to the Lichtenberg High-Performance Computer of the TU Darmstadt. The authors gratefully acknowledge P. Wochner and other beamline scientists at Karlsruhe for the help with the synchrotron XRD measurements, the insightful discussion with W. Sigle, support during TEM sample preparation by U. Salzberger, M. Kelsch and B. Fenk, and the TEM support by Y. Wang, K. Hahn and P. Kopold.

**Conflict of interest**

The authors declare no conflict of interest

**References**

[1]    R. Ramesh, D. G. Schlom, *Nat. Rev. Mater.* **2019**, *4*, 257.

[2]    H. Boschker, J. Mannhart, *Annu. Rev. Condens. Matter Phys.* **2017**, *8*, 145-164

[3]    H. Y. Hwang, Y. Iwasa, M. Kawasaki, B. Keimer, N. Nagaosa, Y. Tokura, *Nat. Mater.* **2012**, *11*, 103.

[4]    J. H. Ngai, F. J. Walker, C. H. Ahn, *Annu. Rev. Mater. Res.* **2014**, *44*, 1.

[5]    S. R. Spurgeon, P. V. Balachandran, D. M. Kepaptsoglou, A. R. Damodaran, J. Karthik, S. Nejati, L. Jones, H. Ambaye, V. Lauter, Q. M. Ramasse, K. K. S. Lau, L. W. Martin, J. M. Rondinelli, M. L. Taheri, *Nat. Commun.* **2015**, *6*, 6735.

[6]    W. Wang, M. W. Daniels, Z. Liao, Y. Zhao, J. Wang, G. Koster, G. Rijnders, C. Z. Chang, D. Xiao, W. Wu, *Nat. Mater.* **2019**, *18*, 1054.

[7]     D. Pesquera, G. Herranz, A. Barla, E. Pellegrin, F. Bondino, E. Magnano, F. Sanchez, J. Fontcuberta, *Nat. Commun.* **2012**, *3*, 1189.

[8]     D. Yi, J. Liu, S. L. Hsu, L. Zhang, Y. Choi, J. W. Kim, Z. Chen, J. D. Clarkson, C. R. Serrao, E. Arenholz, P. J. Ryan, H. Xu, R. J. Birgeneau, R. Ramesh, *Proc. Natl. Acad. Sci. U. S. A.* **2016**, *113*, 6397.

[9]     E. K. Ko, J. Mun, H. G. Lee, J. Kim, J. Song, S. H. Chang, T. H. Kim, S. B. Chung, M. Kim, L. Wang, T. W. Noh, *Adv. Funct. Mater.* **2020**, *30*, 2001486.

[10]    J. Wang, *Annu. Rev. Mater. Res.* **2019**, *49*, 361.

[11]    Z. Li, S. Shen, Z. Tian, K. Hwangbo, M. Wang, Y. Wang, F. M. Bartram, L. He, Y. Lyu, Y. Dong, G. Wan, H. Li, N. Lu, J. Zang, H. Zhou, E. Arenholz, Q. He, L. Yang, W. Luo, P. Yu, *Nat. Commun.* **2020**, *11*, 184.

[12]    H. Liu, Y. Dong, D. Xu, E. Karapetrova, S. Lee, L. Stan, P. Zapol, H. Zhou, D. D. Fong, *Adv. Mater.* **2018**, *30*, e1804775.

[13]    S. Koohfar, A. B. Georgescu, A. N. Penn, J. M. LeBeau, E. Arenholz, D. P. Kumah, *npj Quantum Mater.* **2019**, *4*, 1.

[14]    A. O. Adeyeye, N. Singh, *J. Phys. D: Appl. Phys.* **2008**, *41*, 153001.

[15]    J. I. Martin, J. Nogues, K. Liu, J. L. Vicent, I. K. Schuller, *J. Magn. Magn. Mater.* **2003**, *256*, 449.

[16]    J. Wenisch, C. Gould, L. Ebel, J. Storz, K. Pappert, M. J. Schmidt, C. Kumpf, G. Schmidt, K. Brunner, L. W. Molenkamp, *Phys. Rev. Lett*. **2007**, *99*, 077201.

[17]    J. Mannhart, H. Boschker, T. Kopp, R. Valenti, *Rep. Prog. Phys.* **2016**, *79*, 084508.

[18]    R. Ashoori, Nature **1996**, *379*, 413.

[19]    W. A. Tisdale, X. Y. Zhu, *Proc. Natl. Acad. Sci. U. S. A.* **2011**, *108*, 965.

[20]    C. B. Rong, D. Li, V. Nandwana, N. Poudyal, Y. Ding, Z. L. Wang, H. Zeng, J. P. Liu, *Adv. Mater.* **2006**, *18*, 2984.

[21]    G. Laskin, H. Wang, H. Boschker, W. Braun, V. Srot, P. A. van Aken, J. Mannhart, *Nano Lett.* **2019**, *19*, 1131.

[22]    P. F. Chen, Z. Huang, M. S. Li, X. J. Yu, X. H. Wu, C. J. Li, N. N. Bao, S. W. Zeng, P. Yang, L. L. Qu, J. S. Chen, J. Ding, S. J. Pennycook, W. B. Wu, T. V. Venkatesan, A. Ariando, G. M. Chow, *Adv. Funct. Mater.* **2020**, *30*, 1909536.

[23]    D. Kan, R. Aso, R. Sato, M. Haruta, H. Kurata, Y. Shimakawa, *Nat. Mater.* **2016**, *15*,

432.

[24] J.-S. Lee, D. Arena, P. Yu, C. Nelson, R. Fan, C. Kinane, S. Langridge, M. Rossell, R. Ramesh, C.-C. Kao, *Nat. Mater.* **2010**, *105*, 257204.

[25] G. Koster, L. Klein, W. Siemons, G. Rijnders, J. S. Dodge, C. B. Eom, D. H. A. Blank, M. R. Beasley, *Rev. Mod. Phys.* **2012**, *84*, 253.

[26] X. D. Wu, S. R. Foltyn, R. C. Dye, Y. Coulter, R. E. Muenchausen, *Appl. Phys. Lett.* **1993**, *62*, 2434.

[27] W. L. Lu, P. Yang, W. D. Song, G. M. Chow, J. S. Chen, *Phys. Rev. B* **2013**, *88*, 214115.

[28] H. Jeong, S. G. Jeong, A. Y. Mohamed, M. Lee, W. S. Noh, Y. Kim, J. S. Bae, W. S. Choi, D. Y. Cho, *Appl. Phys. Lett.* **2019**, *115*, 092906.

[29] H. Boschker, T. Harada, T. Asaba, R. Ashoori, A. V. Boris, H. Hilgenkamp, C. R. Hughes, M. E. Holtz, L. Li, D. A. Muller, H. Nair, P. Reith, X. R. Wang, D. G. Schlom, A. Soukiassian, J. Mannhart, *Phys. Rev. X.* **2019**, *9*, 011027.

[30] D. Kan, R. Aso, H. Kurata, Y. Shimakawa, *Adv. Funct. Mater.* **2013**, *23*, 1129.

[31] S. Kolesnik, Y. Z. Yoo, O. Chmaissem, B. Dabrowski, T. Maxwell, C. W. Kimball, A. P. Genis, *J. Appl. Phys.* **2006**, 99, 08F501.

[32] A. F. Marshall, L. Klein, J. S. Dodge, C. H. Ahn, J. W. Reiner, L. Mieville, L. Antagonazza, A. Kapitulnik, T. H. Geballe, M. R. Beasley, *J. Appl. Phys.* **1999**, 85, 4131.

[33] L. Klein, J. S. Dodge, C. H. Ahn, J. W. Reiner, L. Mieville, T. H. Geballe, M. R. Beasley, A. Kapitulnik, *J. Phys. Condens. Matter* **1996**, 8, 10111.

[34] Y. Kats, I. Genish, L. Klein, J. W. Reiner, M. R. Beasley, *Phys. Rev. B* **2005**, 71, 100403.

[35] S. Kang, Y. Tseng, B. H. Kim, S. Yun, B. Sohn, B. Kim, D. McNally, E. Paris, C. H. Kim, C. Kim, T. W. Noh, S. Ishihara, T. Schmitt, J.-G. Park, *Phys. Rev. B* **2019**, *99*, 045113.

[36] Y. J. Chang, C. H. Kim, S.-H. Phark, Y. Kim, J. Yu, T. Noh, *Phys. Rev. Lett.* **2009**, *103*, 057201.

[37] D. Ruzmetov, Y. Seo, L. J. Belenky, D. M. Kim, X. L. Ke, H. P. Sun, V. Chandrasekhar, C. B. Eom, M. S. Rzchowski, X. Q. Pan, *Adv. Mater.* **2005**, *17*, 2869.

[38] J.-P. Maria, H. L. McKinstry, S. Trolier-McKinstry, *Appl. Phys. Lett.* **2000**, *76*, 3382.

[39] A. Vailionis, H. Boschker, W. Siemons, E. P. Houwman, D. H. A. Blank, G. Rijnders, G. Koster, *Phys. Rev. B* **2011**, *83*, 064101.

[40] H.-T. Jeng, S.-H. Lin, C.-S. Hsue, *Phys. Rev. Lett.* **2006**, *97*, 067002.

[41]  I. Solovyev, P. Dederichs, V. J. P. R. B. Anisimov, *Phys. Rev. B* **1994**, *50*, 16861.

[42]  A. Huang, S.-H. Hung, H.-T. Jeng, *Appl. Sci.* **2018**, *8*, 2151.

[43]  M. T. Johnson, P. J. H. Bloemen, F. J. A. denBroeder, J. J. deVries, *Rep. Prog. Phys.* **1996**, *59*, 1409.

[44]  D. Yi, H. Amari, P. P. Balakrishnan, C. Klewe, A. T. N'Diaye, P. Shafer, N. Browning, Y. Suzuki, *Phys. Rev. Appl.* **2021**, *15*, 024001.

[45]  A. Vailionis, H. Boschker, Z. Liao, J. R. A. Smit, G. Rijnders, M. Huijben, G. Koster, *Appl. Phys. Lett.* **2014**, *105*, 131906.

[46]  S. Agrestini, Z. Hu, C.-Y. Kuo, M. Haverkort, K.-T. Ko, N. Hollmann, Q. Liu, E. Pellegrin, M. Valvidares, Herrero-Martin, *Phys. Rev. B* **2015**, *91*, 075127.

[47]  L. Klein, Y. Kats, A. Marshall, J. Reiner, T. Geballe, M. Beasley, A. Kapitulnik, *Phys. Rev. Lett.* **2000**, *84*, 6090.

[48]  H. Wang, V. Srot, B. Fenk, G. Laskin, J. Mannhart, P. A. van Aken, Micron **2021**, *140*, 102979.

[49]  M. Hytch, F. Houdellier, F. Hue, E. Snoeck, Nature **2008**, *453*, 1086.

[50]  A. De Backer, K. H. W. van den Bos, W. Van den Broek, J. Sijbers, S. Van Aert, Ultramicroscopy **2016**, *171*, 104.

[51]  Y. Wang, U. Salzberger, W. Sigle, Y. Eren Suyolcu, P. A. van Aken, Ultramicroscopy **2016**, *168*, 46.

[52]  H. Wang, X. Jiang; Y. Wang, R. W. Stark, P. A. van Aken, J. Mannhart, H. Boschker, Nano letters **2020**, 20, 88.

[53]  Y. A. Abramov, V. G. Tsirelson, V. E. Zavodnik, S. A. Ivanov, I. D. Brown, *Acta. Crystallogr. B. Struct. Sci. Cryst. Eng. Mater.* **1995**, *51*, 942.

[54]  S. L. Dudarev, G. A. Botton, S. Y. Savrasov, C. J. Humphreys, A. P. Sutton, *Phys. Rev. B* **1998**, *57*, 1505.

# Supporting information

**Tunable Magnetic Anisotropy in Patterned SrRuO$_3$ Quantum Structures: Competition between Lattice Anisotropy and Oxygen Octahedral Rotation**

*Hongguang Wang[*], Gennadii Laskin, Weiwei He, Hans Boschker, Min Yi, Jochen Mannhart &, and Peter A. van Aken*

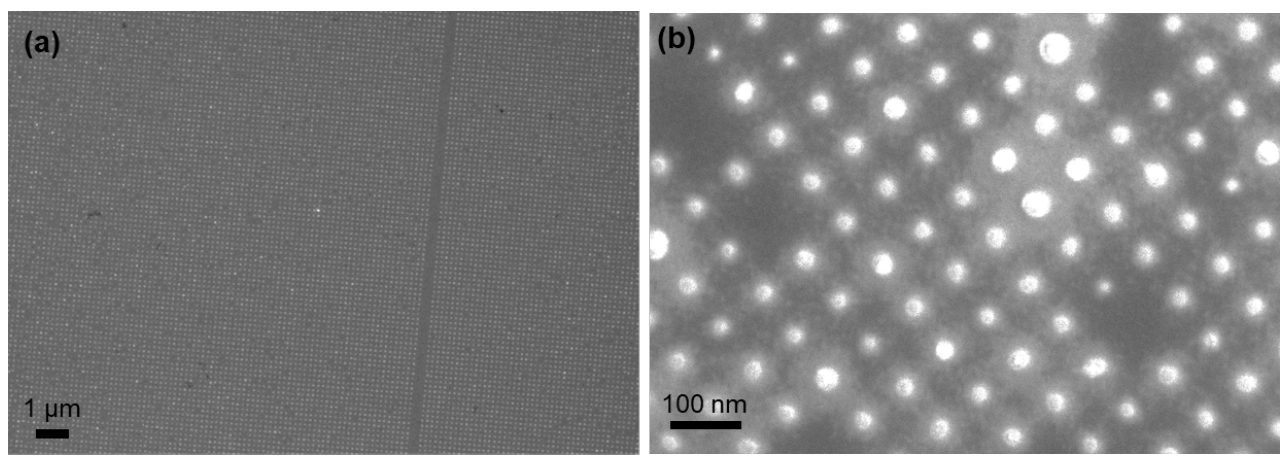

Figure S1. SEM image of the array of 20 nm-sized (a) and 15-20 nm-sized (b) nanodots. Both the number of missing nanodots and the size inhomogeneity increase with smaller nanodots sizes.

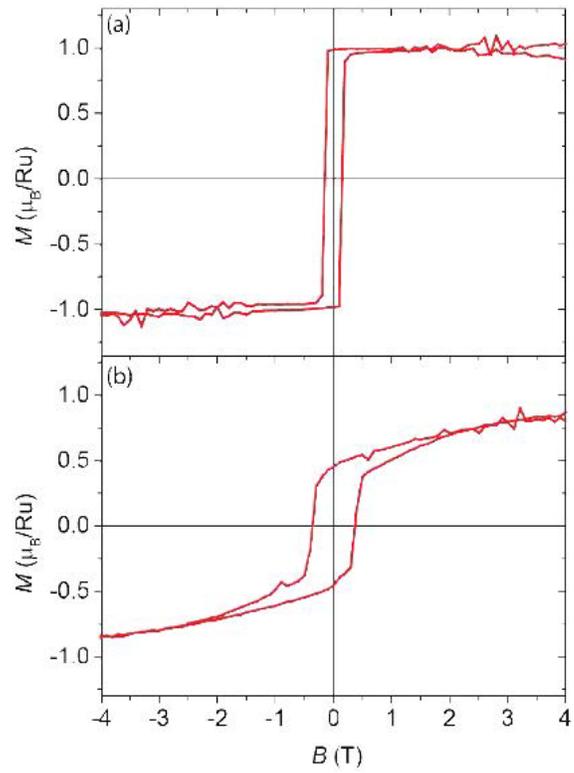

Figure S2. Hysteresis loops of a SrRuO$_3$ film grown on (100) SrTiO$_3$ measured at $T$ = 15K in the (a) out-of-plane and (b) in-plane oriented magnetic field.

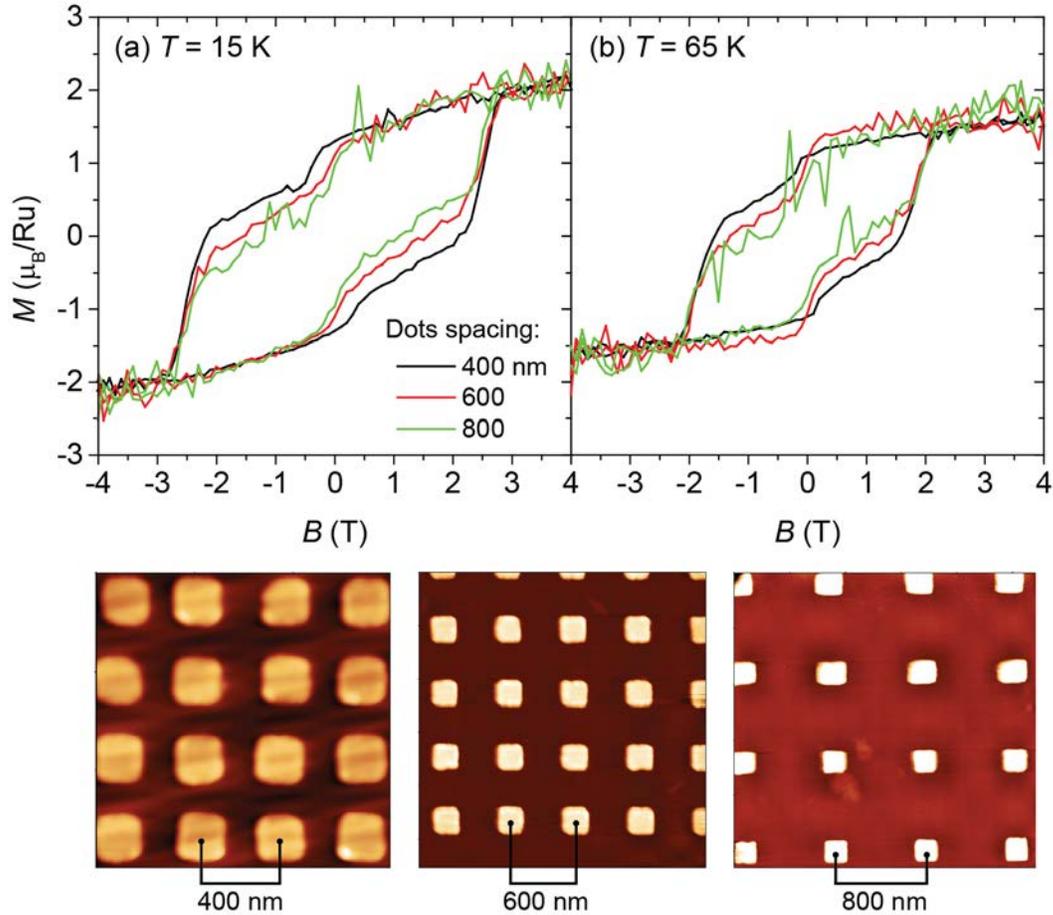

**Figure S3.** Magnetic hysteresis loops measured on rectangular arrays of SrRuO$_3$ dots with dot spacings of 400, 600, and 800 nm. The size of a dot is 200 nm and the thickness is 12 nm. The curves are measured at (a) T = 15K and (b) T = 65 K. The bottom part of the figure depicts AFM images of the corresponding samples.

To study the effects of the interaction between the dots, a series of samples has been produced. The samples contain square-shaped nanodots of SrRuO$_3$ with a side length of 200 nm and a thickness of 12 nm formed into rectangular arrays with different periodicities of 400, 600, and 800 nm. The increasing dot spacing is intended to decrease the amount of the stray field experienced by the dots, changing the setup from a strongly interacting one to an isolated ensemble. A further reduction of the interactions is challenging as samples become more diluted with increasing periodicity, thus decreasing the signal-to-noise ratio in the magnetic measurements. Figure S3 shows hysteresis loops measured for three samples with different dot spacing. All samples were measured with an out-of-plane orientated magnetic field at temperatures of T = 15K (Fig. S3(a)) and T = 65K (Fig. S3(b)). At high fields, above ca. 2.5 T, the loops for different samples coincide quite well, while the behavior is discrepant at a smaller field. The major differences are observed in the low magnetic field region – from 0 to 2.5T – in one of the shoulders of the loops. All samples show the same magnetization at the beginning of each scan (the starting point is at a high positive magnetic field of 5-7 T). When the field is swept downwards, the magnetization follows essentially the same line for samples with different dot periodicities. While sweeping the field down, close to 0 T, the behavior changes and the magnetization of every sample pursues its own path. At a field of 2-3T in the opposite direction, the curves merge again and the magnetization saturates rather rapidly. In the region of changing behavior, the smaller dot density results in a smaller total magnetization. The more dilute samples tend to reduce the amount of the external stray field by forming some complex magnetic structures. Such structures can be magnetization states with no or reduced external field at remanence, e.g., magnetic vortices, closure domains states, or demagnetization "checkerboard" structures formed by dots ensembles[1]. From theoretical considerations, these magnetic structures can result in a behavior similar to the described above. Thus, we suggest that the magnetic interactions within and between nanodots are likely to explain the observed steps at low fields.

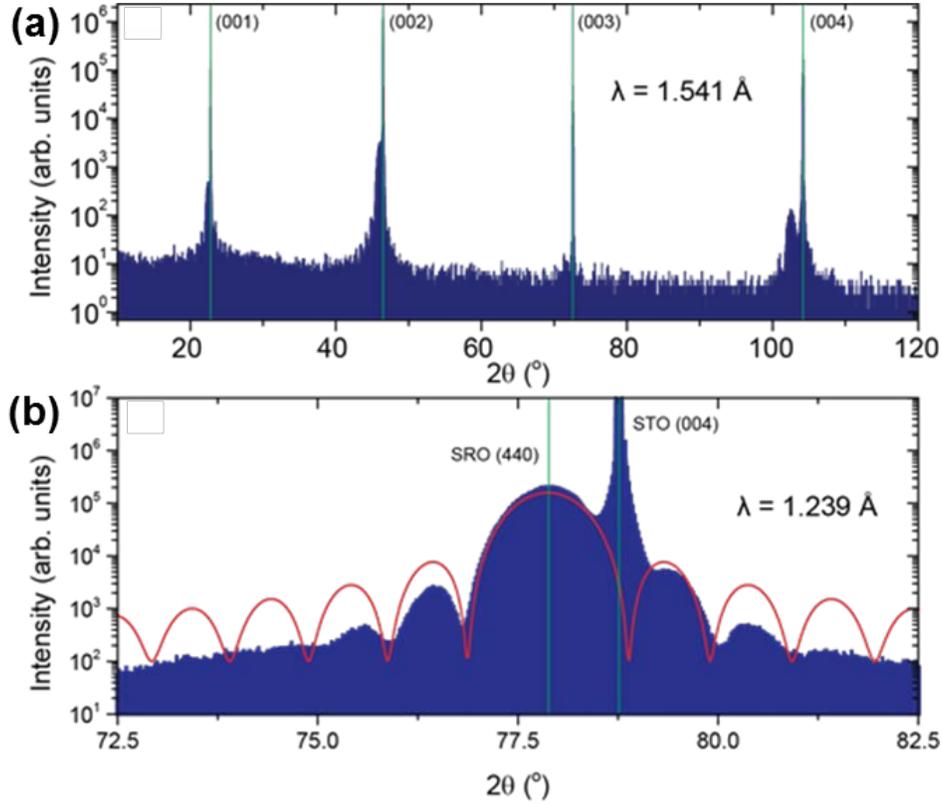

**Figure S4.** (a) A large-scale $\theta/2\theta$ scan of a 10-nm-thick SRO film grown on STO (100) performed with a wavelength $\lambda = 1.541$ Å. The peaks correspond to the diffraction maxima of a cubic crystal with the lattice periodicity of 3.905 Å matching the lattice parameter of STO. The positions of SRO peaks are close to the left of the corresponding STO ones. (b) $\theta/2\theta$ scan close to the (004) STO peak performed at the ANKA synchrotron facility ($\lambda = 1.239$ Å). Green lines mark peak positions, and the red curve depicts the calculated Laue oscillations. The SRO peak is shifted by about 0.9° towards smaller angles, indicating a larger out-of-plane lattice parameter of SRO. That fits the out-of-plane lattice elongation caused by the Poisson effect and the biaxial in-plane compressive strain applied to the film from the substrate. The broadening of SRO peaks results from the small thickness of the thin film. The thickness and the out-of-plane lattice parameter of the film can be extracted from the Laue intensity oscillations.[2] The calculated thickness and out-of-plane lattice parameters are 10 nm and 3.943 Å, respectively.

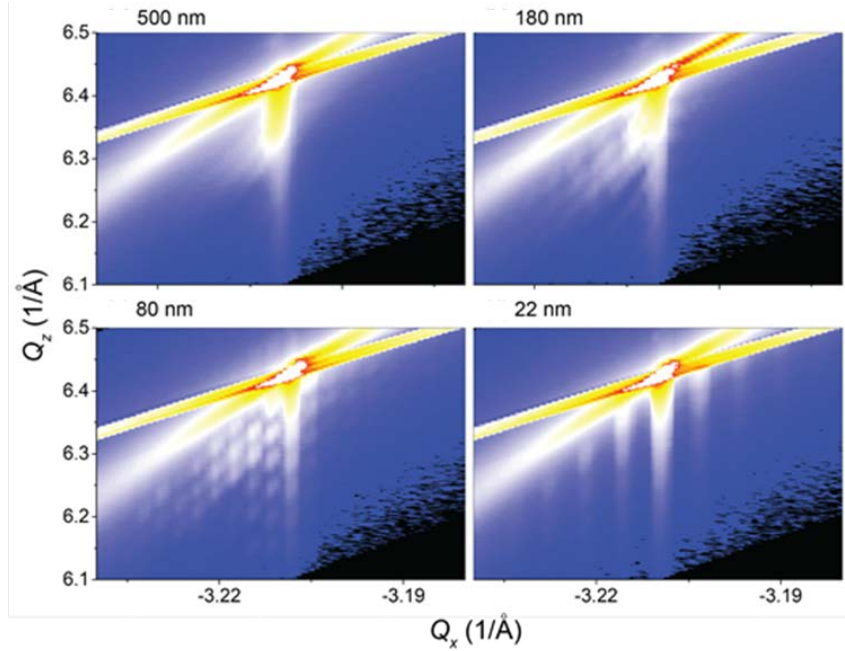

**Figure S5**. Reciprocal space maps close to the (204) peak of STO for arrays of SRO nanodots of different diameters: 500 nm, 180 nm, 80 nm, 22 nm.

The diffraction on the large-scale structure of the arrays corresponding to different nanodot periodicities can be observed. RSMs for nanodot arrays show an elongated SRO peak located under the STO one. In addition, depending on the array periodicity, additional diffraction peaks appear on the sides of the main one, which originates from XRD on the large-scale structure of the arrays. The distance between the peaks in reciprocal space increases with decreasing real space periodicity of nanodot arrays. This distance describes the correct value of the nanodot spacing in the arrays. Due to the small amount of SRO in these nanostructures, the counting statistics of the XRD measurement are low. Also, the appearance of additional maxima has the downside that the intensity of the radiation gets distributed between all subpeaks, further reducing the counting statistics of the measurement. Furthermore, the epitaxial strain in SRO nanodots is gradually relaxed from the SRO/STO interface to the SRO surface, and the absolute value of strain relaxation is small depending on the size of SRO nanodots (only 0.7 % on the surface layer of 30 nm-sized nanodots). Therefore, the lattice changes may be too small to be observed in averaged weak XRD signal, and a solid conclusion about the structure of the patterned arrays cannot be drawn according to these XRD data.

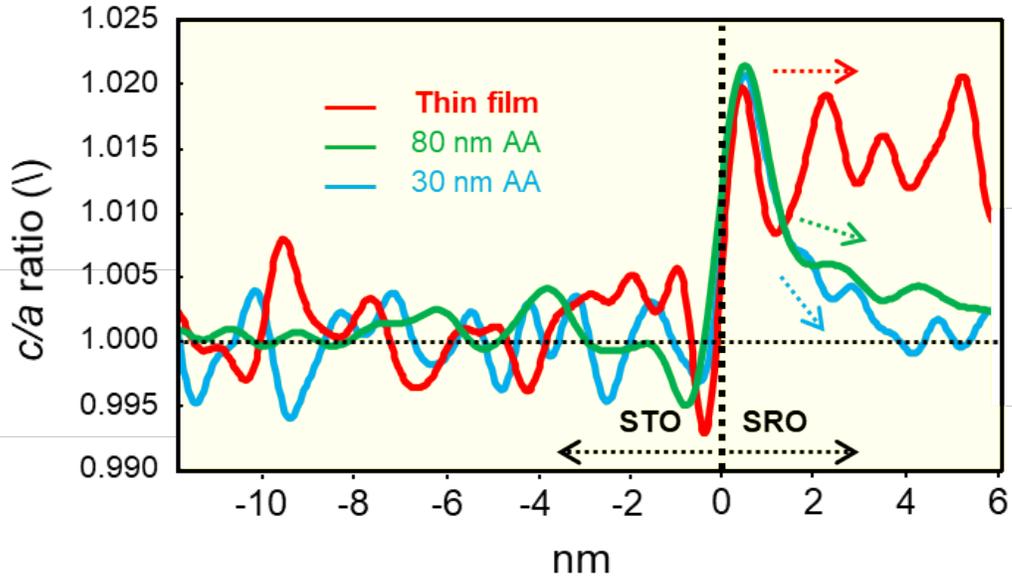

**Figure S6.** The vertically averaged *c/a* values of the box regions in Figures 3(a), 3(b), and 3(c).

The *c/a* value is identical in SRO thin film. However, the *c/a* value for the SRO part of nanodots attains the maximum close to the interface and gradually drops as moving away from the interface toward the surface. The smaller the nanodot size, the faster the *c/a* value decrease. It is obvious that the averaged *c/a* value in the SRO layer becomes smaller as the size of nanodots reduces.

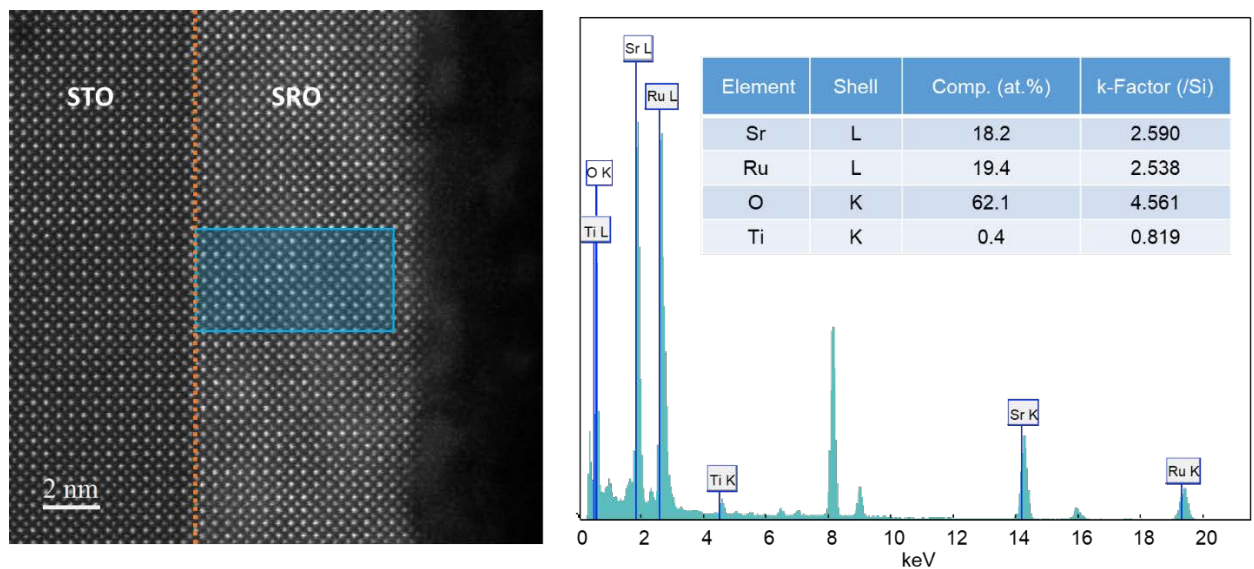

**Figure S7.** EDS spectrum of SRO thin film and corresponding quantification results.

The EDS spectrum has been acquired from the region within the blue box of the SRO thin film. With this spectrum, the atomic ratio of existing elements was then analyzed using the "standardless" quantitative analysis procedure in Gatan DigitalMicrograph. It is noted that, the accuracy of the EDS quantification is under debate, the accuracy of standardless analysis could be 5% ~ 20 % for the relative error [3] [4]. EDS quantifications results demonstrate that the epitaxial SRO thin film is close to the stoichiometry of $SrRuO_3$. A weak Ti signal can be detected due to a redeposition during the ion milling process when the corresponding TEM specimen is prepared.

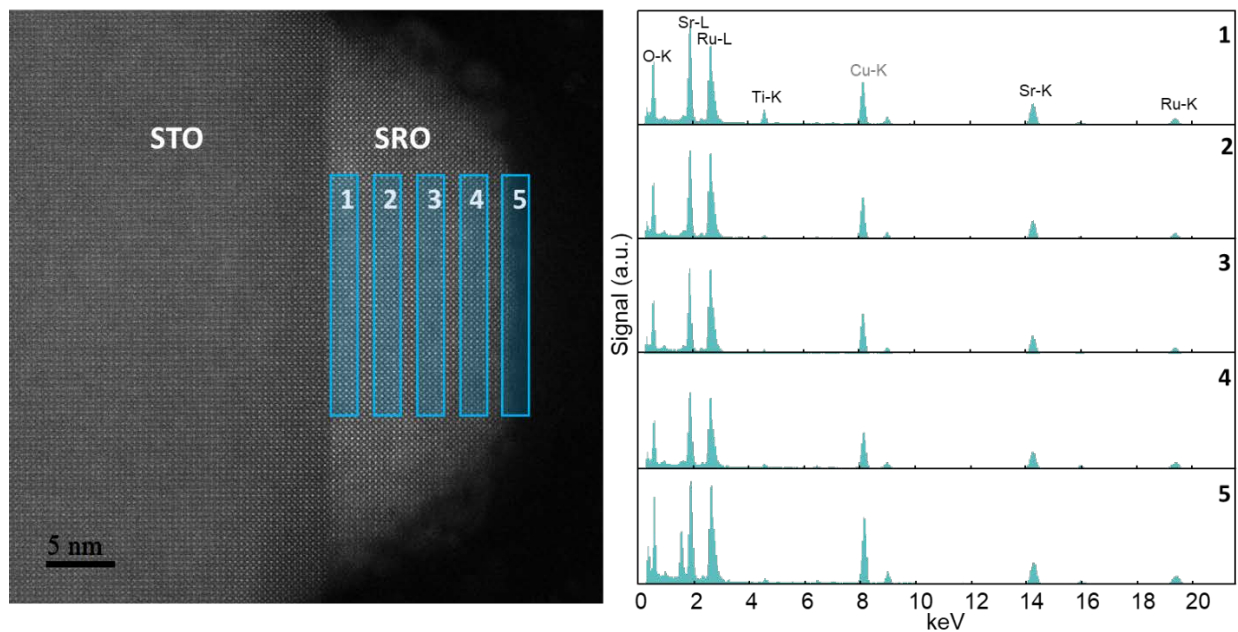

| Regions | Sr (at.%) | Ru (at.%) | O (at.%) | Ti (at.%) |
|---|---|---|---|---|
| 1 | 17.7 | 16.1 | 65.1 | 1.1 |
| 2 | 17.5 | 19.4 | 62.8 | 0.3 |
| 3 | 17.6 | 19.7 | 62.5 | 0.2 |
| 4 | 17.6 | 18.9 | 63.2 | 0.2 |
| 5 | 14.0 | 15.1 | 70.8 | 0.1 |

**Figure S8.** EDS studies of the 30 nm-sized nanodot. EDS spectra of 5 different regions have been acquired and quantified with the same procedure as applied to the spectrum of the SRO thin film. Quantification results for different regions are presented.

We have extracted the EDS signal of 5 different regions of a 30 nm-sized nanodot. Region 1 is close to the interface, regions 2, 3, 4 are in the bulk part of the nanodot, region 5 is at the surface of the nanodot. Compared to the quantification results of the SRO thin film, considering the precision of EDS quantification mentioned above, no apparent changes of the composition of the SRO layer can be observed after nanopatterning. Moreover, the composition throughout patterned SRO nanodots is maintained except for slight deviations close to the surface of SRO and the SRO/STO interface.

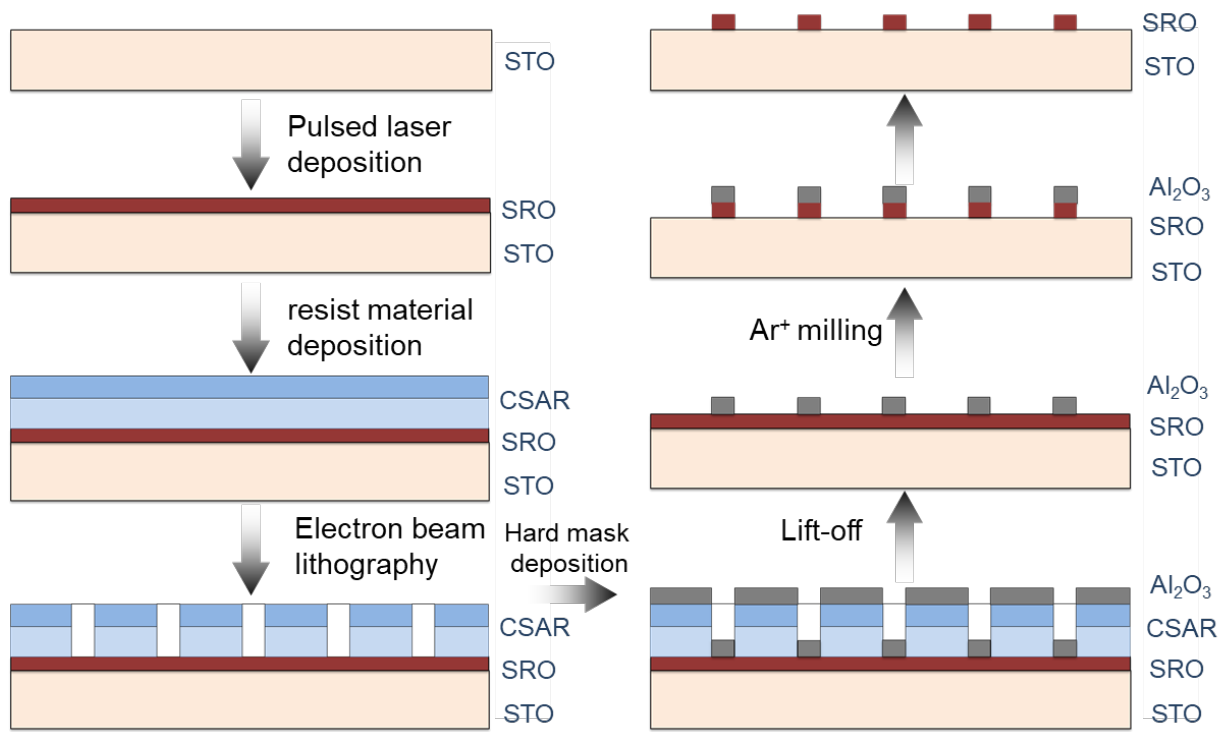

**Figure S9.** Step-by-step fabrication of a SRO nanodot array.

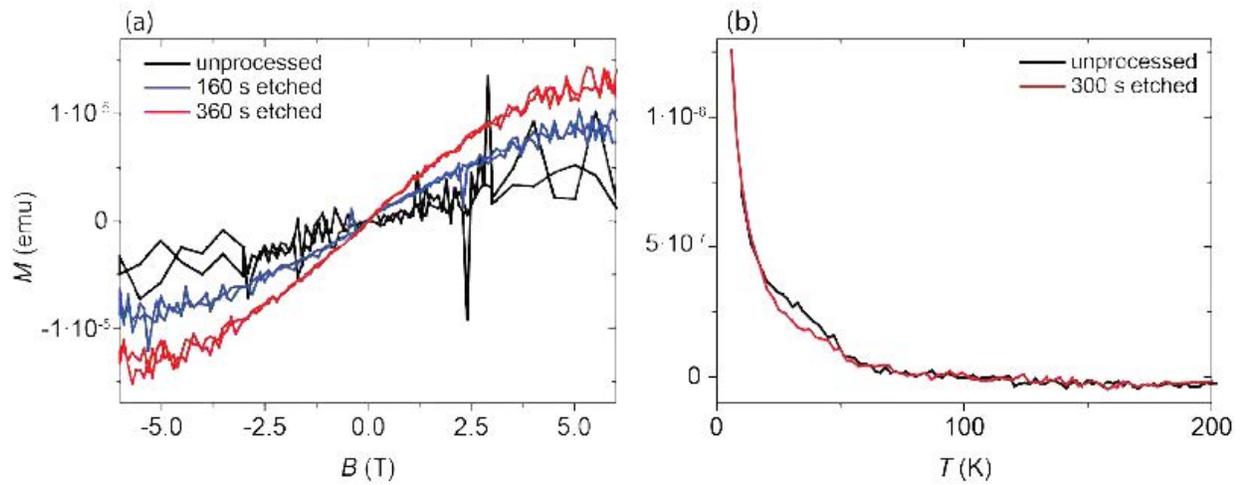

**Figure S10** SQUID magnetometry measurement on bare $SrTiO_3$ substrates before and after the ion irradiation. (a) M(B) curves measured at 15K for a substrate etched for 160 and 360 s. (b) M(T) curves measured at 0.1T for a substrate before and after 300 s etching. All curves show a paramagnetic signal.

References

[1]     L. Landau, J. W. Reiner, L. Klein, *J. Appl. Phys.* **2012**, *111*, 07B901.
[2]      I. K. Robinson, *Phys. Rev. B* **1986**, *33*, 3830.
[3]     D. E. Newbury, *J. Mater. Sci.* **2015**, *50*, 493.
[4]     D. B. Williams,  C. B. Carter, Transmission electron microscopy, Springer, Boston, MA 1996.